\documentclass[12pt,a4paper]{article}
\usepackage[pdfstartview=FitH,colorlinks=true,linkcolor=blue,anchorcolor=black,citecolor=black,urlcolor=blue]{hyperref}
\usepackage{amsmath,amssymb,titling,authblk}
\usepackage{slashed}
\usepackage{amsmath}
\usepackage{amscd}
\usepackage[normalem]{ulem}
\usepackage{appendix}
\usepackage{bbold}
\usepackage{bm}

\usepackage{pdflscape}
\usepackage{yfonts}
\usepackage{amscd}
\usepackage{epsfig}
 \usepackage{microtype}

\usepackage{cite}

\allowdisplaybreaks[4]

\usepackage {color}
\usepackage{braket}
\usepackage{bbm}

\advance\hoffset by -1.1truecm
\advance\voffset by -1.0truecm
\newcommand{\be}{\begin{equation}}
\newcommand{\ee}{\end{equation}}
\newcommand{\bea}{\begin{eqnarray}}
\newcommand{\eea}{\end{eqnarray}}

\newcommand{\del}{\partial}
\newcommand{\dd}{\mathrm d}
\newcommand{\ii}{\mathrm i}
\newcommand{\e}{\mathrm e}
\usepackage{mathrsfs}
\usepackage{authblk}
\usepackage{float}
\restylefloat{figure}
\newcounter{appendice}

\begin{document}

\setlength{\droptitle}{-6pc}

\title{Localizability in $\kappa$-Minkowski Spacetime\thanks{Based on a talk given by FL at the XXVth edition of the conference in Vietri \textsl{PAFT 2019, Current problems  in theoretical physics}, to appear in the special issue dedicated to it.}}

\renewcommand\Affilfont{\itshape}
\setlength{\affilsep}{1.5em}

\author[1,2,3]{Fedele Lizzi\thanks{fedele.lizzi@na.infn.it}}
\author[1,2]{Mattia Manfredonia\thanks{mattia.manfredonia@na.infn.it}}
\author[1,2]{Flavio Mercati\thanks{flavio.mercati@gmail.com}}
\affil[1]{Dipartimento di Fisica ``Ettore Pancini'', Universit\`{a} di Napoli {\sl Federico~II}, Napoli, Italy}
\affil[2]{INFN, Sezione di Napoli, Italy}
\affil[3]{Departament de F\'{\i}sica Qu\`antica i Astrof\'{\i}sica and Institut de C\'{\i}encies del Cosmos (ICCUB),
Universitat de Barcelona, Barcelona, Spain}

\date{}

\maketitle

\vspace{-2cm}

\begin{abstract}\noindent
Using the methods of ordinary quantum mechanics we study $\kappa$-Minkowski space as a quantum space described by noncommuting self-adjoint operators, following and enlarging~\cite{Localization}. We see how the role of Fourier transforms is played in this case by Mellin transforms. We briefly discuss the role of transformations and observers.

\end{abstract}

\newpage


The first quantum space has been the phase space of a point particle under quantization. To study it, the tool which usually employed is due to Dirac, and is based on the correspondence of position and momenta with  \emph{noncommuting} operator on the Hilbert space of pure states. The commutation relation is given by
\be
[ \hat q^i, \hat p_j]=\ii\hbar\delta^i_j \ , \label{Heisencommrel}
\ee
where the dimensionful quantity $\hbar$ acts as deformation parameter. This quantum space is extremely well known, and in the following we will recall some of its features as needed.

In this note we want to use similar methods, but apply them to a different space~\cite{kmink1,majid22,koso,koso2,Dimitrijevic2003,
 MicheleKappa,FlavioKappaDifferential,FlavioKappaLightCone,TPJCW1,TPJCW2, Lizzi:2006bu}: $\kappa$-Minkowski spacetime. These methods have been useful for the study of the quantum properties of various models of $\kappa$-Poincar\'e invariant scalar field theories \cite{TPJCW1,TPJCW2}. Besides, the geometric (spectral) properties, \textit{\`a la} Connes, of the $\kappa$-Minkowski spacetime have been investigated in \cite{dandrea,marseille1,marseille2,Matassa}. defined by the following commutation relations in a four dimensional spacetime:
\be
[ x^0, x^i]=\ii\lambda  x^i, \ [ x^i, x^j]=0, \  \label{commrel}
\ee
where $i,j=1,2,3$, $ x^0=c \,t$ with $c$  speed of light, the deformation parameter $\lambda$ has the dimension of a length\footnote{Often the deformation parameter $\lambda$ is indicated by $\frac 1\kappa$, hence the name.}  and time has a scale of $\frac\lambda c$. There exist many versions of $\star$-products which reproduce the commutation relation \eqref{commrel}, see \emph{e.g.} \cite{DurhuusSitarz, Meljanac:2008ud, TPJCW1}. Some of the results collected here have been first present in~\cite{Localization}, where further details can be found.
The $\kappa$-Minkowski space is the homogeneous space of the $\kappa$-Poincar\'e Hopf algebra (quantum group)~\cite{LukierskiInventsKpoincare1,LukierskiInventsKpoincare2,
 ZakrzewskiInventsKPGroup,Lukierski_kappaPoincareanydimension,
 UniquenessOfKappa}. 
$\kappa$-Minkowski is again a quantum space, but this time the deformation is not of a phase space, and accordingly the deformation parameter is not the quantum of action, but a length, which is natural to think of the order of Planck length, although this is not strictly necessary.

Both non trivial commutation relations~\eqref{Heisencommrel} and ~\eqref{commrel} give rise to uncertainty relations, which are however quite different. In phase space we have 
\be
\Delta p \Delta q\geq \frac\hbar2
\ee
while in $\kappa$-Minkowski involves only a relation between the time coordinate and any of the space ones, 
\be
\Delta x^0 \Delta x^i \geq \frac\lambda2 |\langle x^i \rangle| \,, \label{uncertx0xi}
\ee
There is no uncertainty for measurement of space coordinates among themeselves.

We follow this analogy and apply Dirac's correspondence principle to study the kinematics of a spacetime ``quantized'' by $\lambda$ (as phase space was quantized by $\hbar$). In the end of the paper are going to briefly discuss how  measurements of position and time are concerned in such a framework, and which are the states.  One might wonder whether localizability of a state depend on the reference frame or not. As an example, for the quantum phase space of one particle the algebra of positions and momenta is invariant under classical translations and rotations. However, the algebra~\eqref{commrel} is clearly not invariant under the classical action of the Poincar\'e group (in particular under translations and boosts) but it is  under a deformation of the Poincar\'e group. This makes the group manifold itself into a noncommutative space,  as a consequence, different observers will not agree in general on the localizability properties of the same state.
We stress that in the present we will strictly discuss the \emph{kinematics} of systems in $\kappa$-Minkowski but not their dynamics. We are interested in the relation between the  noncommutativity of spacetime  and  the localizability of events, \emph{independently of the dynamical laws they are subject to}. The limits to localizability then could be understood as the effective description of gravitational excitations that intervene when enough energy is concentrated into a small region in order to localize an event~\cite{Doplicher0,CAMead}, and this would indeed be a dynamical effect. This connection with a fundamental quantum field theory of gravity is only crude. At the moment it can be made precise in 2+1 dimensions where relations like~(\ref{commrel}) emerge  once one integrate away the gravitational degrees of freedom in a model of quantum gravity coupled with matter~\cite{FreLiv,MatWel}.

\subsubsection*{Example: Quantum Phase Space for a particle in 3D}
We shortly discuss the archetypical noncommutative geometry of the phase space of a single quantum particle in three dimension. The material is known from elementary courses in quantum mechanics, but we wish to stress a few features which will be used analogously for $\kappa$-Minkowski spacetime. The quantization turns the coordinates $( q^i, p_i)$ of the  six-dimensional phase space into noncommuting operators: $\hat q^i$ and $\hat p^i$. These are commonly represented as operators on the Hilbert space$L^2(\mathbb R^3_q)$ of square integrable functions of space as\footnote{To simplify the notation we indicate by $q$ and $p$ the corresponding three-vectors, avoiding the use of a notation like $\vec q$.}:
\be
\hat q^i\psi(q)=q^i\psi(q) \ ; \qquad \hat p_i\psi(q)=-\ii\hbar\frac\del{\del q^i} \psi(q) \ . \label{pqreponq}
\ee
They  are unbounded selfadjoint operators with a dense domain. The spectrum is the real line (for each $i$). 
The eigenvalue equation
\be
\del_q \psi(q)= \alpha \psi(q) \ , \qquad \alpha \in \mathbbm{C}^3 \label{momeigenfline}
\ee 
admits only improper eigenfunctions $\e^{ \alpha \cdot q }$ with $\alpha=\ii k$ a pure imaginary number. We all know that those are physically interpreted as infinite plane waves of given a frequency.
As usual  $|\psi(q)|^2$ for normalized functions gives the  probability density to find the particle at position $q$. The wave function, being a complex quantity, contains also the information about the density probability of the momentum operator.

If we choose the $\hat p_i$ as the complete set of observable, we have
\be
\hat q^i\phi(p)=\ii\hbar \frac\del{\del p^i}\phi(p) \ ; \qquad \hat p_i\phi(p)=p_i \phi(p) \ . \label{pqreponp}
\ee
Of course,  $\psi(q)$ and $\phi(p)$ carry exactly the same information and are connected by a \emph{Fourier} transform\footnote{It is important that the Fourier transform is an isometry, \emph{i.e.} it maps normalized functions of positions into normalized functions of momenta.}:
\be
\psi(q)=\frac 1{(2\pi)^{\frac32}} \int \dd^3 p\, \phi(p) \e^{\frac\ii\hbar p\cdot q}
\ee
This is of course well known. Let us now consider the case of $\kappa$-Minkowski in the same spirit.

\subsubsection*{Time and position of events in $\kappa$-Minkowski space} \label{se:timeandpositioninkappa}
Let us begin by considering the $\hat x^i$'s as a complete set of observables on the  Hilbert space $L^2(\mathbb R^3_x)$. We will represent\footnote{ The representations of the algebra generated by~\eqref{commrel} are discussed in detail in~\cite{Agostini:2005mf, DabrowkiPiacitelli}} the $\hat x^\mu$ as operators on this space as\footnote{The most general class of operator compatible with the algebra can be found in~\cite{MeljanacStojic}(see also~\cite{MeljanacMercati,MeljanacMercati2})}
\bea\label{Rep_kappa-Minkowski}
\hat x^i \psi(x)&=&x^i\psi(x), \nonumber\\
\hat x^0\psi(x)&=&\ii \lambda \left(\sum_i x^i\del_{x^i} + \frac32\right)\psi(x)=\ii\lambda \left(r\del_r + \frac32\right)\psi(x).
\eea
The $\frac32$ factor is necessary to have symmetric operators. Here, $\hat x^0$ plays the role that $\hat p$ played in quantum phase space. The $\hat x^0$ operator is a dilation operator, and this suggests the use of a polar basis. The polar coordinates $\hat \theta,\hat \varphi$ do not correspond to well defined self-adjoint operators, but if one defines $\hat r \cos \hat \theta=\hat x^3$ and $\hat  r \, \e^{\ii\hat \varphi}=(\hat x^1+\ii \hat x^2)$, then gets
\be
[\hat x^0, \cos \hat \theta]=[\hat x^0,\e^{\ii \hat \varphi}]= 0 \,, \qquad [\hat x^0,\hat r ] = \ii \lambda \hat r \,.
\ee
In fact $\hat x^0$ commutes with all spherical harmonics, or in general functions of $\hat \theta$ and $\hat \varphi$ independent on $r$.  In the following we will consider the vectors of $L^2(\mathbb R^3_x)$ to be functions of the kind $\psi=\sum_{lm}\psi_{lm}(r)Y_{lm}(\theta,\varphi)$. Moreover, since the angular variables commute with everything, we will focus on the radial part only.
The uncertainty relation~\eqref{uncertx0xi} can be expressed in its polar version 
\be
\Delta \hat  x^0 \Delta \hat r \geq \frac\lambda2 |\langle \hat  r \rangle|. \label{uncertx0r}
\ee

The operator $\hat x^0$ is symmetric, but we should verify its self-adjointness domain. Integrating by parts, one finds:
\be
\int \dd r  r^2\, \psi_1^* \ii \lambda\left(r\del_r + \frac32\right) \psi_2= \ii \lambda\int \dd r  r^2\, \psi_1^* \frac32\ \psi_2-\int\dd r\,  \ii\lambda \del_r\left( r^3 \psi_1^*\right) \psi_2 + \psi_1^* r^3 \psi_2 \bigg|^\infty_0 \,.
\ee
One can see that the boundary term vanishes if $\psi_1$ and $\psi_2$ vanish at infinity faster than  $r^{-\frac32}$, which is true for all square-integrable (according to the measure $\int\! \dd r r^2$) functions. In the origin the condition imposed is weaker than the one imposed by square-integrability.
Let us now look for the spectrum and the (improper) eigenvectors. Monomial in $r$ are formal solutions of the eigenvalue problem:
\be
\ii\lambda \left(r\del_r + \frac32\right) r^\alpha=\ii\lambda (\alpha+\frac32) r^\alpha=\lambda_\alpha r^\alpha, \label{eigenfunctionsgeneric}
\ee
therefore eigenvalues are  
\be 
\lambda_\alpha= \ii \lambda(\alpha+\frac32).
\ee
Monomials are never acceptable physical states, vector of $L^2(\mathbbm{R}^3_x)$. They are never square integrable becasue they will always diverge either at the origin or at infinity (remember the correct measure!).  When the real part of $\alpha=-\frac32$ the function diverges both at zero and at infinity. Compare this with plane waves. An exponential, which is the formal solution of the eigenvalue equation for momentum, also diverges either at plus or minus infinity. When the exponential is pure imaginary it diverges at both ends, but only marginally, i.e.\ in a way that it is a good distribution with a large dens domain. In this case also the eigenvalues are real, as it befits to a self-adjoint operator.
This behaviour is mirrored in the $\kappa$-Minkowski case. Eigenfunctions~\eqref{eigenfunctionsgeneric} are acceptable distributions if and only if
\be
\alpha=-\frac32 + \ii \tau , \ \ \ \tau \in \mathbb{R}
\ee
The spectrum of the time operator is real and goes from minus infinity to plus infinity.
What plane waves were for momentum in quantum phase space, now are the distribution  
\be
T_\tau=\frac{r^{-\frac32-\ii\tau}}{\lambda^{-\ii\tau}}=r^{-\frac32}\e^{-\ii\tau\log\left(\frac r\lambda\right)},
\ee
They have a well defined inner product with every vector in the domain of $\hat x^0$. The distribution has the correct dimension of a length to the 3/2, the factor of $\lambda$ is there to avoid taking the logarithm of a dimensional quantity.

The operator $\hat x^0$ is selfadjoint, therefore the $T_\tau$ are a complete base. Both $r, \theta, \varphi$ and $\tau, \theta, \varphi$ are complete sets of observables. Completeness implies that any function of $r$ can be \emph{isometrically} expanded in terms of the $T_\tau$.
\begin{equation}
\psi(r,\theta,\varphi)=\frac1{\sqrt{2\pi}}\int_{-\infty}^{\infty}\dd\tau\, r^{-\frac32}\e^{-\ii\tau\log\left(\frac r\lambda\right)} \widetilde\psi(\tau,\theta,\varphi)=\mathcal{M}^{-1}\left[\widetilde\psi(\tau,\theta,\varphi), \ 
r \right],
\end{equation}
\begin{equation}
\widetilde \psi(\tau,\theta,\varphi)=\frac1{\sqrt{2\pi}}\int_0^\infty \dd r\, r^{\frac12}\e^{\ii\tau\log\left(\frac r\lambda\right)} \psi(r,\theta,\varphi)=\mathcal{M}\left[\psi(r,\theta,\varphi), \-\frac{3}{2}+\ii \tau \right]. \label{Mellintrans}
\end{equation}
The above integrals define a Mellin transform together with its inverse. Both are isometries between square integrable functions of $r$ with measure $\dd r r^2$ and functions of $\tau$. We see that Mellin transform are for $k$-Minkowski what Fourier are for quantum phase space. Hereafter we will often omit the explicit dependence on $\theta$ and $\varphi$ when there is no confusion. 

We are now going to make an important assumption: that the usual theory of measurement holds. Namely, observables are self-adjoint operator, bounded or unbounded. The possible results of a measurement are the points of the spectrum, and when a pure state is expressed in the spectral decomposition given by the complete set of (proper or improper) eigenfunctions of the observables, the modulus square of the function gives the probability density. With these assumptions,  
the average time measured by a particle in the state described by $\psi$ with spherical symmetry gives:
\be
\langle \hat x^0\rangle_\psi=4\pi \int r^2\dd r \overline \psi(r) \ii\lambda\left(r\del_r+\frac32\right) \psi(r) \label{x0average}
\ee
If $\psi$ is real it results $\langle \hat x^0\rangle_\psi=0$. Hence only complex valued functions will have non vanishing  mean value of time (in analogy with quantum phase space, where real functions have a vanishing mean value of the momentum). The probability of measuring a given value of $\tau$  is given by $|\widetilde \psi(\tau)|^2$ for normalised functions.  

To have a feeling for the dimensional quantities consider $\lambda$ to be Planck length. Reinserting dimensionful quantities $\tau=t \frac c\lambda$ with $t=\frac{x^0}c$.  If $t=1$\,s, then $\tau=2\cdot 10^{43}$, an extremely large number. If $t$ is of the order of Planck time, then $\tau\sim 1$.

Various examples of functions can be found in~\cite{Localization}, here we present some further examples of localization of states in space and time. In the following we will consider only the radial coordinate, since, as we said, the angular part does not play any role.
 
\begin{figure}[htb]\center
\includegraphics[width=0.7 \textwidth]{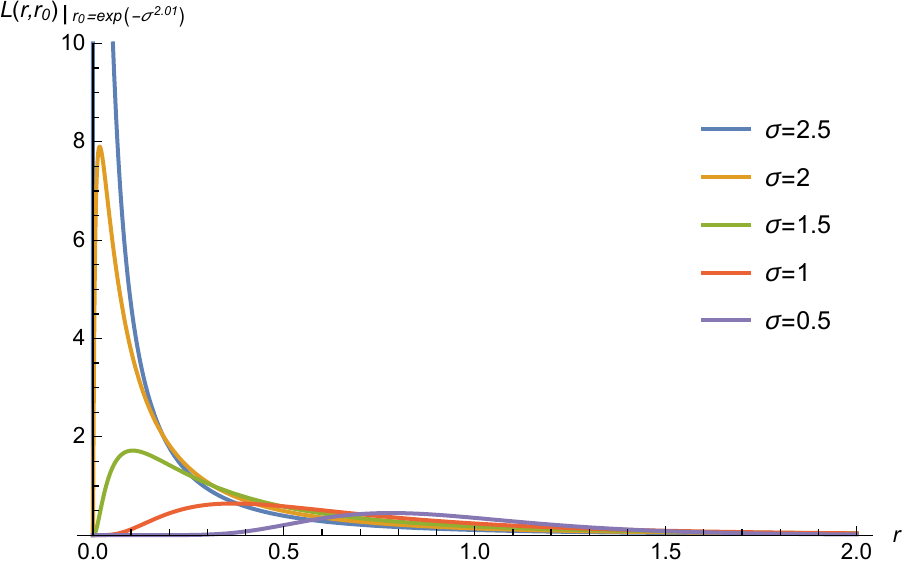}
\caption{\sl The $\sigma \to \infty $ limit of $L(r,r_0)$ when $\xi = e^{-\sigma^{(2+\epsilon)}}$, for $\epsilon = 0.01$. \label{figL}}
\end{figure}

We first note that according to~\eqref{uncertx0r} it is possible to localize a particle both in space and time, as long as the position is the origin. Such a state, being akin to a Dirac's $\delta$, will again be a distribution, but we can present it as limit of ordinary states.
To do so we use functions that saturate the uncertainty bounds provided by \eqref{uncertx0r} (\textit{i.e.} the analogue of what Gaussian coherent states were for quantum phase space). It turns out that this role is played by normalized \emph{log-Gaussians}, plotted in Fig.~\ref{figL},
\be\label{LogGaussians}
L(r,r_0)=N \e^{-\frac{(\log r - \log r_0)^2}{\sigma^2}}=\frac{\e^{-\left(\frac{\log
   \left(\frac{r}{r_0}\right)}\sigma\right)^2}\e^{-
   \frac{9}{16} \sigma^2}}{\sqrt{\sigma} (2 \pi
   )^{3/4} \sqrt{r_0^3}} \,.
\ee
These functions have  a maximum in $r=r_0$, and they localize at $r =r_0$ as $\sigma \to 0$, and  at $r=0$ as $r_0 \to 0$, for any value of  $\sigma \geq 0$ and are plotted in Fig.~\ref{figL}.

\begin{figure}[htb]\center
\includegraphics[width=0.7 \textwidth]{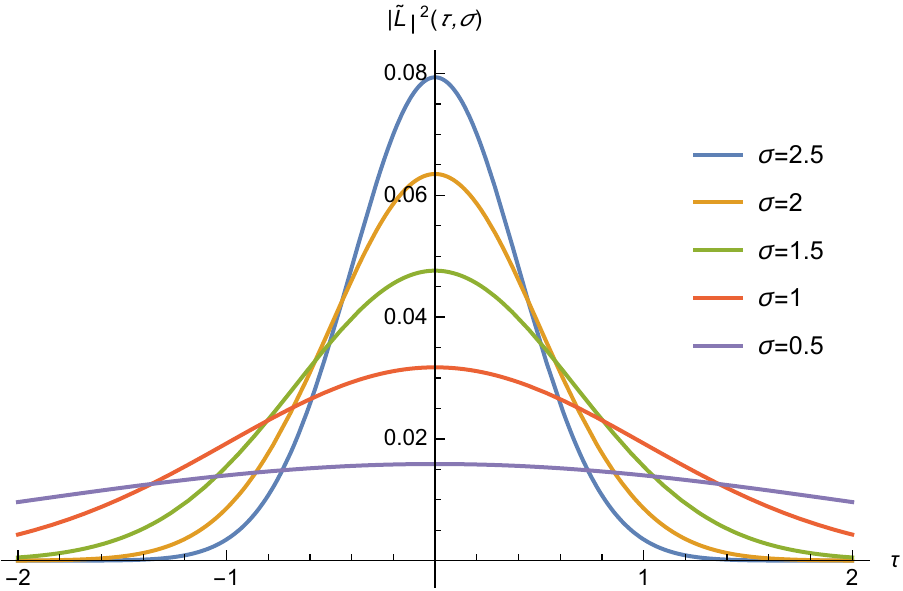}
\caption{\sl The Mellin transforms of the function $L(r,r_0)$ for various values of $\sigma$. \label{figLMellin}}
\end{figure}

The Mellin transform~\eqref{Mellintrans} of these states is a Gaussian:
 \be
\widetilde L(\tau,r_0)=\frac{\sigma^{\frac12}\e^{-\frac{1}{4} \sigma^2\tau 
   (\tau -3 \ii)}  }{2
   \sqrt[4]{2} \pi ^{3/4}}  \left(\frac{r_0}{\lambda}\right)^{\ii \tau } \,,
\ee
With square norm:
\be
\left|\widetilde L(\tau,r_0)\right|^2=\frac{\sigma\e^{-\frac{\sigma^2 \tau
   ^2}{2}}}{4 \sqrt{2} \pi
   ^{3/2}} \,,
\ee
namely, in $\tau$ space the probability density is a Gaussian independent on $r_0$.  The Mellin transforms of the states described in Fig.~\ref{figL} are shown in Fig.~\ref{figLMellin}.

To argue that these states are localized we need to compute both $\langle r^n \rangle_L$ $\langle x_0^n \rangle_L$ on such a state and shows that they all vanish for $r_0 \to 0$.  After some computation we get
\begin{align}
&\label{ExpectationValue_r^n}\langle \hat r^n \rangle_L=\e^{\frac{\sigma^2}8 n(n+6)} r_0^,n\\
&\label{ExpectationValue_x0^n}\langle (\hat x^0)^n \rangle_L = \frac 1 {4 \pi} \left(\frac\lambda\sigma\right)^n \left\{ \begin{array}{lcl} 0 & \ & n\ \mbox{odd}\\
(n-1)!! & \ & n\ \mbox{even}\end{array}\right.
\end{align}
We can see that there is a double limit $r_0\to 0$ and $\sigma\to\infty$ which gives a state which is localized both is space (at $r=0$) and in time. For example, it is sufficient to take $r_0=e^{-\sigma^{2+\epsilon}}$ for any $\epsilon>0$, that all $\langle \hat r^n \rangle_L$ in~(\ref{ExpectationValue_r^n}) and all $\langle (\hat x^0)^n \rangle_L$ in~(\ref{ExpectationValue_x0^n}) go to zero as $\sigma \to \infty$. Although the above example has been done at $\tau=0$ it is possible to shift the state by just multiplying the function by $r^{\ii\tau_0}$. Moreover one can attribute any wavefunction to time while still having the spatial coordinates localized at the origin, just by convoluting this with a function of $\tau$. In other words, we have an  `eigenstate of the origin'\footnote{While we have seen that there is a state corresponding to $\ket o$, there is not a normalized vector corresponding to it. Here (and in the following) we are here performing the usual abuse of notation made when one uses the ket notation $\ket x$ in ordinary quantum mechanics. } $|o\rangle$  completely localized at the origin of spacetime. Furthermore, it can be obtained as a limit of normalized elements of $L^2(\mathbbm{R}^3_x)$. Moreover we have a 1-parameter family of states, $|o_\tau\rangle$ localized at the origin of space at a nonzero time. These states as well can be obtained as limits of normalized elements of $L^2(\mathbbm{R}^3_x)$.

Alternatively one can consider states which are Gaussian in position (not log-Gaussian). The presence of a cusp in the origin does not pose problems since we are not requiring differentiability, at any rate it can be smoothened if necessary. States of the kind 
\be
G(r, r_0,\sigma)=N \e^{\frac{(r-r_0)}{2\sigma}}
\ee
with the normalization
\be
N^{-2}=\frac{1}{2} \sqrt{\pi } \sigma 
   \left(\text{erf}\left(\frac{r_0}{\sigma
   }\right)+1\right)
 \ee
 have tranform
\bea
&\widetilde G(r, r_0,\sigma)=&\nonumber\\&N \frac{\left(\frac{\lambda }{\sigma }\right)^{-\ii \tau}
   \left(\sigma  \Gamma \left(\frac{\ii
   \tau}{2}+\frac{3}{4}\right) \, _1F_1\left(-\frac{\ii
   \tau}{2}-\frac{1}{4};\frac{1}{2};-\frac{r_0^2}
   {\sigma ^2}\right)+2 r_0 \Gamma
   \left(\frac{i t}{2}+\frac{5}{4}\right) \,
   _1F_1\left(\frac{1}{4}-\frac{\ii
   \tau}{2};\frac{3}{2};-\frac{r_0^2}{\sigma
   ^2}\right)\right)}{\sqrt{2} \pi  \sqrt{\sigma }
   \left(\text{erf}\left(\frac{r_0}{\sigma
   }\right)+1\right)}&
\eea
 where $_1F_1$ are hypergeometric functions of the first kind.
 
\begin{figure}[htb]\center
\includegraphics[width=0.7 \textwidth]{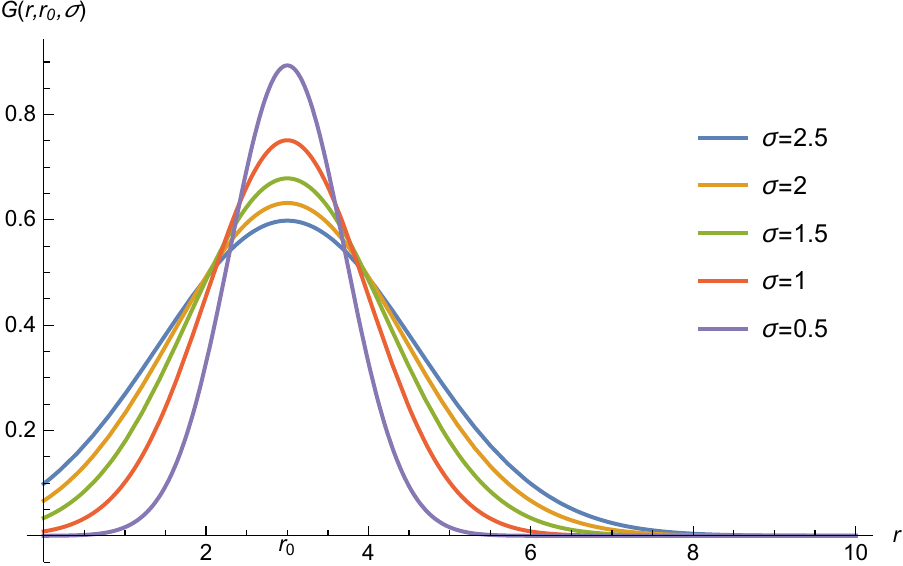}
\caption{\sl The function $G(r,r_0)$ for various values of $\sigma$. \label{figG}}
\end{figure}

\begin{figure}[htb]\center
\includegraphics[width=0.7 \textwidth]{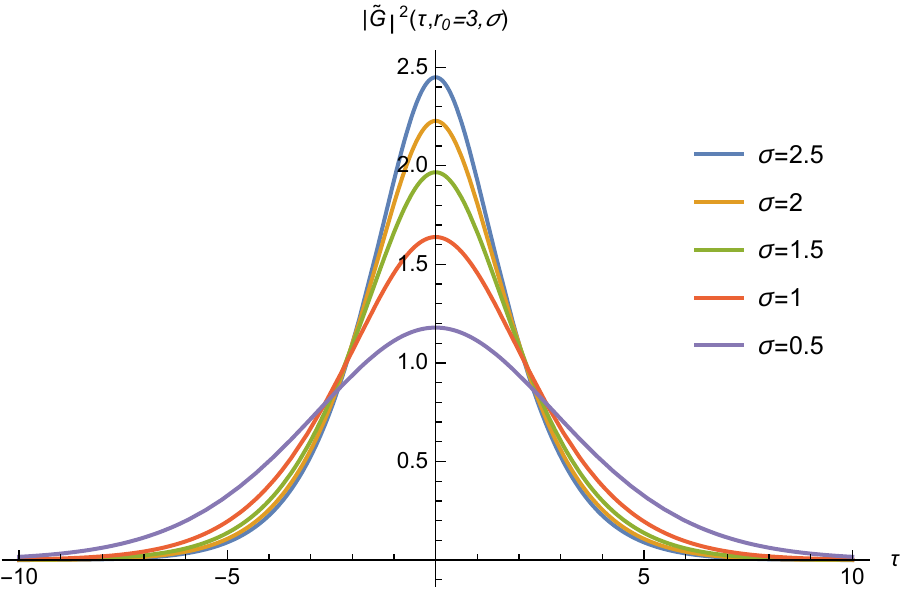}
\caption{\sl The absolute square of Mellin transforms of the function $G(r,r_0)$ for various values of $\sigma$. \label{figGMellin}}
\end{figure}

The analogues of Figs.~\ref{figL} and~\ref{figLMellin} are in Figs.~\ref{figG} and~\ref{figGMellin} respectively.

 Note that in both cases the constant $\lambda$ appears only as a phase, and will not appear in the modulus square of the function. It is nevertheless impossible to set $\lambda=0$ from the onset, as it would give singular functions without a definite limit.

As further example consider a particle localized on a given distance from the origin, regardless of direction. The wave functions in $r$ and $\tau$ spaces are:
\bea
\psi(r)&=&\delta(r-r_0)/r_0^2
\nonumber\\
\widetilde\psi(\tau)&=&\frac1{\sqrt{2\pi}} r_0^{-\frac32} \left(\frac{r_0}\lambda\right)^{\ii\tau}=\frac1{\sqrt{2\pi}} r_0^{-\frac32}\e^{\ii \tau\log \left(\frac{r_0}\lambda\right)} \,, \label{deltatransf}
\eea
The Mellin transfor $\widetilde\psi(\tau)$ is a distrubtion. We can approximate it smearing on a spherical shell
\be
\psi(r)=\left\{\begin{array}{lc}0& r<R_1\\ \sqrt{\frac3{4\pi(R_2^3-R_1^3)}}&R_1\leq r \leq R_2\\ 0 & R_2<r\end{array}\right.
\label{FirstWavefunc}
\ee
its Mellin transform is:
\be
\widetilde\psi(\tau)=\frac1{\sqrt{2\pi}}\sqrt{\frac3{4\pi(R_2^3-R_1^3)}}\left(\frac{R_2^{\frac32+\ii\tau}-R_1^{\frac32+\ii\tau}}{\lambda^{\ii\tau}}\right) \frac2{3+2\ii\tau} \,, \label{FirstWavefuncMellin}
\ee
with probability density:
\be
|\widetilde\psi(\tau)|^2=\frac3{8\pi^2(R_2^3-R_1^3)}\left[ R_2^3+R_1^3-2 R_1^{\frac32}R_2^{\frac32}\cos \left(\tau\log\frac{R_2}{R_1}\right)\right] \frac4{9+4\tau^2}\,,
\label{FirstWavefuncMellinProbDensity}
\ee
which is an even function, which explains why the average value of $\hat x^0$ vanishes. The probability density~(\ref{FirstWavefuncMellinProbDensity}) now is not constant: it is now peaked around $\tau=0$ and it decreases like $\tau^{-2}$ away from the origin.
In the limit $R_1\to R_2$ the Mellin transform~(\ref{FirstWavefuncMellin}) is (up to a constant) the Mellin transform of the delta function,~(\ref{deltatransf}).

Note that in all these examples, when the particle is localised, even if on a spherical shell, the square of the wave function $|\psi (\tau)|^2$ is independent on $\tau$. Localization in space means complete delocalization in time. This is a consequence of the uncertainty, just as phase space the Fourier transform of localised states (Dirac's $\delta$) are plane wave, whose suare modulus does not depend on the momentum. 

\subsubsection*{Is the origin special?}

We have argued that the origin is a special point. Does this mean that somewhere in the universe there is \emph{``the origin''}. A special position in space singled out by the $\kappa$-God? Needless to say, there is no special point, and the space is ``translationally invariant''. The invariance is however not under the usual Poincar\'e group, but under the quantum symmetry called $\kappa$-Poincar\'e. This issue is described in detail in~\cite{Localization}. Here we will give a shortened version.

Implicitly in our discussion, when we were referring to states we were assuming the existence of an observer measuring the localisation of states. This observer is located at the origin, and he can measure with absolute precision where he is. For him ``here'' and ``now'' make sense. He cannot localize with precision states away from him, as a consequence of the noncommutativity of $\kappa$-Minkowski.
What about other observers? A different observer will be in general Poincar\'e transformed, i.e.\ translated, rotated and boosted. These operations are usually performed with an element of the Poincar\'e group.

One motivation to introduce  $\kappa$-Minkowski is its relations to the quantum group $\kappa$-Poincar\'e. We should therefore use {deformed} transformations. We require invariance under the transformation $x^\mu \to  x'^\mu  = \Lambda^\mu{}_\nu \otimes x^\nu + a^\mu \otimes 1$. The commutations relations in a particular basis~\cite{ZakrzewskiInventsKPGroup, FlavioKappaLightCone} are:
\bea
&\left[ a^\mu , a^\nu \right] =  \ii\lambda \left( \delta^\mu{}_0 \, a^\nu - \delta^\nu{}_0 \, a^\mu \right) \,, \qquad [ \Lambda^\mu{}_\nu ,  \Lambda^\rho{}_\sigma ] = 0 &\nonumber\\
&[ \Lambda^\mu{}_\nu , a^\rho ] =  \ii\lambda \left[  \left( \Lambda^\mu{}_\sigma \delta^\sigma{}_0 - \delta^\mu{}_0 \right) \Lambda^\rho{}_\nu + \left( \Lambda^\sigma{}_\nu \delta^0{}_\sigma - \delta^0{}_\nu \right) \eta^{\mu\rho} \right]&
\eea
With coproduct, antipode and counit
\bea
\Delta(a^\mu) = a^\nu \otimes \Lambda^\mu{}_\nu +1 \otimes a^\mu &,&
 \Delta ( \Lambda^\mu{}_\nu ) = \Lambda^\mu{}_\rho \otimes \Lambda^\rho{}_\nu \,,\nonumber\\
S(a^\mu) = - a^\nu  (\Lambda^{-1})^\mu{}_\nu &,&
S ( \Lambda^\mu{}_\nu ) = (\Lambda^{-1})^\mu{}_\nu \,,
\varepsilon(a^\mu) = 0 \,,
\
\varepsilon ( \Lambda^\mu{}_\nu ) = \delta^\mu{}_\nu 
\eea

We represented the $\kappa$-Minkowski algebra as operators. But in doing so we had implicitly chosen an observer.
In order to take into account the fact that there are different observers we enlarge the the algebra (and consequently the space) to include the parameters of the new observers. We call then new set of states as $\mathcal P_\kappa$. Observers in this space cannot prescind from quantization, there are not ``classical'' observers in $\kappa$-Minkowski.
Our (generalized) Hilbert space will now comprise not only function on spacetime (either functions of $r$ or $\tau$), but also functions of the $a$'s and $\Lambda$'s.

We can represent the $\kappa$-Poincar\'e group translations faithfully as
\bea
a^\rho &=&   - \ii \,  \frac\lambda2 \left[  \left( \Lambda^\mu{}_\sigma \delta^\sigma{}_0 - \delta^\mu{}_0 \right) \Lambda^\rho{}_\nu + \left( \Lambda^\sigma{}_\nu \delta^0{}_\sigma - \delta^0{}_\nu \right) \eta^{\mu\rho} \right] \Lambda^\nu{}_\alpha  \frac{\partial}{\partial \omega^\mu{}_\alpha}
\nonumber\\&&+\ii\frac\lambda2 \left( \delta^\rho{}_0 \, q^i \frac{\partial}{\partial q^i} + \delta^\mu{}_i \, q^i \right) + \frac12\mbox{h.c.}
\eea
while the $\Lambda$'s  are simply represented as multiplicative operators.

Therefore that, like spacetime, the space of observers is also noncommutative, but the noncommutativity is only present in the translation sector. This the observers counterpart of the fact that the angular variables commute with everything. Observers which differ only for a rotation will agree on everything. Boosted observes will not, since a boost involves time, which is a quantized variable.
Let us further explore the space of observers, seen as states. First consider the observer located at the origin, which is reached via the identity transformation.
Define  $\ket{o}_{\mathcal P}$ of  with the property:
\be
{}_{\mathcal P}\!\bra{o} f(a,\Lambda) | o \rangle_{\mathcal P} = \varepsilon(f) \,,
\ee
with $f(a,\Lambda)$ a generic  noncommutative function of translations and Lorentz transformation matrices, and $\varepsilon$ the counit.
 This state describes  the Poincar\'e transformation between two coincident observers. The state is  such that all combined uncertainties vanish. Coincident observers are therefore a well-defined concept in $\kappa$-Minkowski spacetime.
A change of observer will transform $x^\mu \to  x'^\mu  = \Lambda^\mu{}_\nu \otimes x^\nu + a^\mu \otimes 1$, primed and unprimed coordinates correspond to different observers.
Identifying $x$ with $\mathbb 1\otimes x$ we generate an extended algebra $\mathcal P\otimes \mathcal M$ which extends $\kappa$-Minkowski by the $\kappa$-Poincar\'e group algebra. 
This algebra takes into account position states and observables. It is a noncommutative algebra, as it befits to noncommutative observers.
\emph{Just as we cannot sharply localize position states, neither we can sharply localize where the observer is.}

We can build the action of the position, translation and Lorentz transformations operator on generic functions of all those variables.
Let us  just state a result, as an example, sending to~\cite{Localization} for further details.
Poincar\'e transforming the origin state $|o\rangle$ by  wavefunction $| g \rangle$ in the representation of   $a^\mu, \Lambda^\mu{}_\nu$ we obtain  for all polynomials in  the transformed coordinates 
\be
x'^\mu = a^\mu \otimes 1 + \Lambda^\mu{}_\nu \otimes x^\nu
\ee
the same expectation value as the one assigned by $|g \rangle$ to the corresponding polynomials in $a^\mu$. In other words, the state of $x'^\mu$ is identical to the state of $a^\mu$.
Uncertainty $\Delta x'^\mu$ is introduced by the uncertainty of the $a^\mu$ close a noncommutative algebra, we cannot know, with absolute precision {where} the new observer is.
 
\subsubsection*{Challenges}

Here we discussed purely the kinematics of $\kappa$-Minkowski. To discuss dynamics, at a quantum level, we would have to introduce phase space, and momentum.   The latter is a most interesting object in this case, the space of momenta is curved~\cite{KappaDeSitterMomentumSpace, MicheleCurvedMomentum, kappaRelativeLocality}. Although the noncommutativity of translations, and the nontrivial curvature of momenta are both aspect of the quantum nature of $\kappa$-Minkowski and $\kappa$-Poincar\'e,  it is not immediate to see momenta as generator of translations in an operatorial way. The connection is subtle and some aspects are being investigate in~\cite{NewMomentum}. Since the observers are indetermined, as well as the states, there is the issue of the possibility of defining in a meaningful way the cluster decomposition principle, or at least meaningful Pauli-Jordsn functions in this context~\cite{MercatiSergolaPauliJordan}. A noncommutative geometry has, by nature, some degree of nonlocality, but is necessary to understand in which way the scale and the details of this nonlocality can be made compatible with quantum field theory.

\subsection*{Acknowledgements}
FL and MM acknowledge the support of  the INFN Iniziativa Specifica GeoSymQFT; FL the Spanish MINECO under project MDM-2014-0369 of ICCUB (Unidad de Excelencia `Maria de Maeztu'): FM the Action CA18108 QG-MM from the European Cooperation in Science and Technology (COST) and partial support from the Foundational Questions Institute (FQXi).

\end{document}